\documentclass{article}

\PassOptionsToPackage{numbers, compress}{natbib}

    \usepackage[final]{neurips_2023}

\usepackage[utf8]{inputenc} 
\usepackage[T1]{fontenc}    
\usepackage{hyperref}       
\usepackage{url}            
\usepackage{booktabs}       
\usepackage{amsfonts}       
\usepackage{nicefrac}       
\usepackage{microtype}      
\usepackage{xcolor}         
\usepackage{graphicx}
\usepackage{amssymb}
\usepackage{mathtools}
\usepackage{amsthm}
\usepackage{bm}
\usepackage{cleveref}
\usepackage{dsfont}
\usepackage{algorithm}
\usepackage{algpseudocode}
\usepackage{multirow}
\usepackage{pifont}
\usepackage{arydshln}

\hypersetup{hidelinks}
\def\1{\bm{1}}
\def\G{\mathcal{G}}
\def\V{\mathcal{V}}
\def\E{\mathcal{E}}

\def\x{\bm{x}}

\newcommand{\mask}{\texttt{[MASK]}}
\newcommand{\purity}{\mathfrak{p}}
\newcommand{\trans}{\bm{Q}}
\newcommand{\cat}{\text{Cat}}
\newcommand{\seq}{\bm{s}}
\newcommand{\res}{s}

\newcommand{\struct}{\bm{x}}
\newcommand{\vocab}{\mathcal{V}}
\newcommand{\length}{L}

\newcommand{\cmark}{\ding{51}}%
\newcommand{\xmark}{\ding{55}}%

\title{Fast non-autoregressive inverse folding\\  with discrete diffusion}

\author{%
  John J. Yang \\
  Massachusetts Institute of Technology\\
  \texttt{johnyang@mit.edu} \\
  \And
  Jason Yim \\
  Massachusetts Institute of Technology\\
  \texttt{jyim@csail.mit.edu}
  \And
  Regina Barzilay \\
  Massachusetts Institute of Technology\\
  \texttt{regina@csail.mit.edu}
  \And
  Tommi Jaakkola \\
  Massachusetts Institute of Technology\\
  \texttt{tommi@csail.mit.edu}
}

\begin{document}

\maketitle

\begin{abstract}
Generating protein sequences that fold into a intended 3D structure is a fundamental step in \textit{de novo} protein design.
De facto methods utilize autoregressive generation, but this eschews higher order interactions that could be exploited to improve inference speed.
We describe a non-autoregressive alternative that performs inference using a constant number of calls resulting in a \textit{23 times speed up} without a loss in performance on the CATH benchmark.
Conditioned on the 3D structure, we fine-tune ProteinMPNN to perform discrete diffusion with a purity prior over the index sampling order.
Our approach gives the flexibility in trading off inference speed and accuracy by modulating the diffusion speed. Code: \url{ https://github.com/johnyang101/pmpnndiff}
\end{abstract}

\section{Introduction}
\label{sec:introduction}
\textit{De novo} protein design aims to design proteins from first principles without modifying an existing protein \citep{huang2016coming}.
This involves designing protein 3D structures for a desired function such as binding then determining the sequence that would fold \textit{in-vivo} into the designed structure.
The first step is the protein structure generation problem for which RFdiffusion has become state-of-the-art \citep{watson2023novo}.
The aim of this work is to develop a discrete diffusion model towards the second step of sampling sequences conditioned to fold into a desired 3D structure, referred to as \textit{inverse (protein) folding} \citep{hsu2022learning}.

Generative models have already become the preferred tool for inverse folding: ProteinMPNN \citep{proteinmpnn} is a widely used method with successful experimental validation.
However, ProteinMPNN has two limitations.
The first is its autoregressive decoding which scales linearly and can be prohibitively slow for large proteins.
The second is the uniformly random decoding order which is likely suboptimal given substantial evidence of higher-order interactions occurring in protein evolution \citep{starr2016epistasis}.

\begin{figure*}[!ht]
\begin{center}
\includegraphics[width=\textwidth]{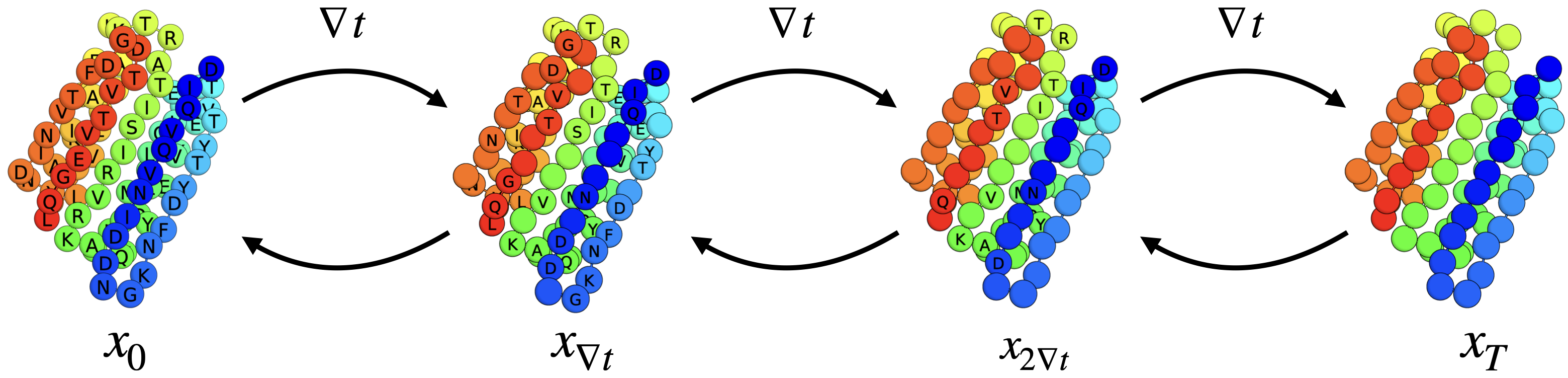}
\caption{Overview. Starting from a sequence and structure, the forward process masks residues on each step.
The reverse process unmasks on each step given only the structure and all residues initializd to \mask.
Our framework allows for flexible decoding strategies by \textit{striding} over intervals of time $\nabla t$ in each reverse step, i.e. the diagram uses 3 steps to decode the whole sequence.
}
\label{fig:absorbing}
\end{center}
\end{figure*}

Inspired by RFdiffusion, in which a pre-trained protein folding model is fine-tuned with diffusion, we use a pre-trained ProteinMPNN and fine-tune it with diffusion.
We explore multiple variants of discrete diffusion and find the best configuration to result in equivalent performance as ProteinMPNN on foldability while using 23 times less compute.
Our best diffusion model is depicted in \Cref{fig:absorbing}.
Our contribution enables exciting possibilities to extend ProteinMPNN using conditional diffusion models for improvement in both speed and controllable generation.

The paper is structured as follows. Related work is provided in \Cref{sec:related_work}.
\Cref{sec:method} describes our method of fine-tuning ProteinMPNN with discrete diffusion.
\Cref{sec:experiments} evaluates each diffusion variant on the CATH benchmark \citep{orengo1997cath}.
We focus on analyzing tradeoffs in diversity, speed, and designability of sequence designs.
\Cref{sec:discussion} concludes with limitations and future directions.


\section{Method}
\label{sec:method}
We first provide background on the problem formulation and ProteinMPNN in \cref{sec:background}.
Next, \cref{sec:diffusion} describes discrete diffusion and its variants we implement for fine-tuning ProteinMPNN.

\subsection{Background}
\label{sec:background}
\textbf{Problem formulation.} The task of inverse folding is to design a sequence $\seq \in \vocab^\length$ that folds into a given backbone structure $\struct \in \mathbb{R}^{\length \times 4 \times 3}$ where $\vocab$ is the vocabulary of 20 amino acids plus a mask token \mask and $\length$ is the number of residues.
Superscripts $\seq^{(t)}$ will be used to refer to time $t \in [1,\dots,T]$ while subscripts $\res_i$ refer to residues $i$ (i.e. index).
Here, $\struct$ refers to the atomic coordinates of the 5 canonical backbone atoms $\{N, C_\alpha, C, C_\beta, O\}$.
It has been well-documented that sequences with as low as 30\% sequence similarity can fold into the same structure \cite{krissinel2007relationship}.
Therefore, the goal is to learn a distribution of possible sequences that fold into a structure, $p(\seq|\struct)$.

\textbf{ProteinMPNN.}
First described in \citet{proteinmpnn}, we provide a brief summary.
Each residue is represented as a node in an attributed graph $\G(\x) = (\V(\x), \E(\x))$.
$\V(\x)$ encodes geometric features such as orientation and sequence index of each residue; $\E(\x)$ constructs a $k$-nearest neighbor graph based on $C_\alpha$ distance and encodes relative geometric features.
The neural network uses message passing to learn embeddings of protein geometry through a masked language modeling objective that was first described in \citet{Ingraham2019}.
Inference is performed by sampling a uniformly random decoding order then autoregressively decoding residues one by one.


\subsection{Discrete diffusion}
\label{sec:diffusion}
We provide a overview of discrete diffusion models (D3PM) described in \citet{austin2021structured}.
Discrete diffusion models are a class of latent variable generative models that are defined by a fixed forward Markov process $q(\seq^{(1:T)}|\seq^{(0)})=\prod_{t=1}^Tq(\seq^{(t)}|\seq^{(t-1)})q(\seq^{(0)})$ from a starting sequence $\seq^{(0)}$ and sequence of increasingly noisy latent variables $\seq^{(1:T)}=(\seq^{(1)},\dots,\seq^{(T)})$.
The goal is to learn a model that parameterizes the reverse process $p_\theta(\seq^{(0:T)}, \struct) = \prod_{t=1}^T p_\theta(\seq^{(t-1)} | \seq^{(t)})$ where,
\begin{align}
    p_\theta(\seq^{(t-1)} | \seq^{(t)}) = \sum_{\seq^{(0)}} q(\seq^{(t-1)} | \seq^{(t)}, \seq^{(0)}) p_\theta(\seq^{(0)} | \seq^{(t)}).
    \label{eq:marginal}
\end{align}
Thus, a model with weights $\theta$ learns how to \textit{denoise} by predicting $p_\theta(\seq^{(0)} | \seq^{(t)})$.
During training, we optimize the evidence-weighted lower bound (ELBO) between $q(\seq^{(t-1)} | \seq^{(t)}, \seq^{(0)})$ and $p(\seq_{t-1} | \seq_{t})$. For absorbing state diffusion, the ELBO reduces to the cross-entropy loss over masked residues, 
\begin{align}\label{eq:loss}
    \mathcal{L} = \mathbb{E}_{q(\seq^{(0)})} \left[
            \sum_{t=1}^T \frac{T - t + 1}{T}  \mathbb{E}_{q(\seq^{(t)} | \seq^{(0)})} \left[
                \sum_{i=1}^L \mathds{1}\{\res^{(t)}_i = \mask\}
                \log{p_\theta\left(\hat{\res}^{(0)}_i | \seq^{(t)}\right)}
            \right]
        \right].
\end{align}
Following \citet{bondtaylor2021unleashing}, each inner expectation is re-weighted to take into account the difficulty of denoising steps.
Discrete diffusion depends on the noising process using a categorical distribution over each residue $q(\res^{(t)}_i | \res^{(t-1)}_i) = \cat(\res^{(t)}_i; \bm{p}=\res^{(t-1)}_i\trans_t)$ that is parameterized by the probabilities $\bm{p}$ of each category where $\trans_t$ is a doubly stochastic transition matrix with $\res_i^{(t-1)}$ represented as a one-hot encoding.
Each residue is noised independently.
We can derive a closed form for sampling the $t$-step marginal and reverse step,
\begin{align}
    q(\res_i^{(t)} | \res_i^{(0)}) &= \cat\left( \res^{(t)}_i; \bm{p}=\res^{(0)}_i\overline{\trans}_t\right), \quad \text{where} \quad \overline{\trans}_j=\prod_{j=1}^t\trans_j \label{eq:marginal}\\
    q(\res^{(t-1)}_i | \res^{(t)}_i, \res_i^{(0)}) &= \cat\left(\res^{(t-1)}_i ; \bm{p}=\frac{\res^{(t)}_i\trans_t^\top \odot \res^{(0)}_i\overline{\trans}_{t-1}}{\res^{(0)}_i \overline{\trans}_{t} (\res^{(t)}_i)^\top} \right).
\end{align}


We choose to use the \textbf{absorbing state} transition matrix, 
\begin{align}\label{eq:trans_t}
\trans_t = (1 - \beta_t)\mathds{I} + \beta_t \mathds{1}e_m^\top
\end{align}
where $\beta_t$ controls the masking rate, the vocabulary is extended with a mask token, and $e_m$ is a vector with 1 on the index of the mask token and 0 elsewhere.
On each forward step, a percentage of the unmasked tokens transition to mask and stay masked until the final step, where all tokens are masked.
The reverse process performs the opposite with unmasking starting with residue as \mask.

\subsection{Diffusion improvements to ProteinMPNN}
\label{sec:pmpnn}

ProteinMPNN's autoregressive sampling can be seen as an absorbing state diffusion that unmasks one residue on each reverse step with a uniformly random decoding order.
Given this perspective, we sought to extend ProteinMPNN to be more efficient with discrete diffusion.
In the context of inverse folding, diffusion is conditioned on the structure, $p_\theta(\seq_{t-1} | \seq_t, \struct)$.
Borrowing techniques from \citet{bondtaylor2021unleashing}, we first fine-tune ProteinMPNN with non-autoregressive diffusion training and strided sampling.
Second, we utilize more informed sampling orders based on a purity prior \citep{tang2023improved}.

\paragraph{Non-autoregressive diffusion.} 
We fine-tune ProteinMPNN with absorbing state diffusion where a subset of the residues are masked/unmasked on each step.
We choose a linear schedule for masking where, for a length $L$ protein, approximately $\lfloor \frac{t}{T}L\rfloor$ will be masked at time $t$. 
The $t$-step marginal probability $\bm{p}=\res_i^{(0)} \overline{\trans}_t$ in \cref{eq:marginal} is then,
\begin{equation}
\res_i^{(0)} \overline{\trans}_t = \prod_{j=1}^{t} (1-\beta_j) \res^{(0)}_i + \prod_{j=1}^{t} \beta_j \mathds{1}e_m^\top = \left(1 - \frac{t}{T}\right) \res^{(0)}_i + \frac{t}{T} \mathds{1}e_m^\top \label{eq:pmpnn_marginal}
\end{equation}
where $\beta_t$ is set such that $t/T = \prod_{j=1}^t\beta_j$.
Training proceeds using \cref{eq:marginal} to noise and $\mathcal{L}$ \cref{eq:loss} as the loss.
To sample, all residues are initially masked then $\lfloor \frac{1}{T}L\rfloor$ residues are unmasked on each step according to a decoding order (to be discussed soon).
Using $T$ steps during sampling can still be prohibitively slow.
We explored efficient sampling where on each step we decode $\lfloor \frac{\nabla t}{T}L\rfloor$ residues and only require $\lceil \frac{\nabla t}{T}\rceil$ steps where $\nabla t$ is an integer value representing the \textit{stride interval}.
A depiction of strided sampling is in \Cref{fig:absorbing}.

\paragraph{Purity Prior.}
Since structure is correlated with evolutionary couplings \citep{hopf2014sequence}, we sought to bias the masking/unmasking order based on couplings strengths, but our preliminary attempts were not successful.
Instead, we utilize a purity prior to bias the order. The \textit{purity} of residue index $i \in \{1,\dots,\length\}$ at step $t$ is defined as 
\begin{align}\label{eq: purity}
    \purity(i, t) = \max_{j \in \{1,\dots,\length\}}p_\theta((\seq_0)_i=j | \seq_t)
\end{align}
Purity can be thought of as the model's confidence in its prediction at a given index relative to the other indices.
We hypothesized ProteinMPNN would be more confident about jointly predicting coupled residues.
On each forward/reverse step, we perform importance sampling based on $\purity$ to determine the next location to mask/unmask (see \citet{tang2023improved}).

\section{Experiments}
\label{sec:experiments}
To evaluate ProteinMPNN diffusion, we report results on the CATH 4.2 Single Chain benchmark previously reported in ProteinMPNN.

\paragraph{Metrics.} Following ProteinMPNN, we sample 8 sequences for each structure in the CATH test set and calculate the following metrics.
The most common metric is \textit{sequence recovery}, defined as percentage of correct amino acids relative to the native sequence, as the primary metrics \citep{hsu2022learning,gao2022pifold,yi2023graph}.
\textit{Designability} is defined as the RMSD error of a pre-trained protein folding model (we use ESMFold \citep{Lin2022ESMFold}) to predict the intended structure from the sampled sequence.
Multiple works have found designability to correlate with experimental success \citep{proteinmpnn,watson2023novo, wicky2022hallucinating}, such evidence does not exist for sequence recovery.
Therefore, we designate designability as the main metric to optimize.

In addition, we report \textit{diversity} as the average pairwise Levenshtein distance between sequences for a structure.
Lastly, \textit{speed} is the average wall clock time (in seconds) to sample one sequence per protein in the test set using a single NVIDIA V100.

\paragraph{Diffusion fine-tuning.} 
We set $T=100$ and use $\nabla t = 20$ when using strided sampling for a total of 5 model calls to sample any sequence.
Depending on the protein length, each model call will have variable wall clock time. 
Dropout is set to 0.1 during training. The Adam optimizer is initialized with a learning rate of $10^{-4}$.
We first pre-train ProteinMPNN with the standard next-token prediction objective as done in \citet{proteinmpnn} on the CATH single chain training set for 200 epochs with batch size of 10,000 tokens.
ProteinMPNN is then fine tuned using the diffusion objective (\Cref{eq:loss}) for 1000 epochs with a batch size of 5k tokens. Models were trained on NVIDIA RTX A6000-48GB.

    
    
\subsection{Results}
Our results are presented in \Cref{table:results}.
Using the same model, we evaluate three different sampling variants based on whether purity order and strided sampling $(\nabla t > 1)$ are used.
Our baseline does not include purity or strided sampling.
Using purity to determine the decoding order results in improvement over recovery and designability but a drop in diversity.
Including strided sampling gives a 10 times improvement in speed.
In comparison to ProteinMPNN, we find sequence recovery is worse with diffusion but designability remains on par with a slight drop in diversity.
The advantage with diffusion is clear: a 23 times speed up with a small drop in diversity and designability.

\begin{table}[ht]
\centering
\caption{Performance of discrete diffusion variants.}
\begin{tabular}{lll| cccc}
\toprule
& Purity & $\nabla t$ & Recovery ($\uparrow$) & Diversity ($\uparrow$) & Designability ($\downarrow$) & Speed ($\downarrow$)\\
\toprule
\multirow{3}{*}{Diffusion} & \xmark & 1 & 33.6 \% & 0.624 & 2.855 & 355.7\\
 & \cmark & 1 & 39.6 \% & 0.424 & 2.158 & 329.8\\
 & \cmark & 20 & 39.9 \% & 0.420  & 2.112 & 33.7\\
\hdashline
ProteinMPNN & & & 47.9 \% & 0.386 & 2.007 & 768.3\\
\bottomrule
\end{tabular}
\vspace{3pt}

\label{table:results}
\end{table}

\section{Discussion}
\label{sec:discussion}
The development of efficient and accurate methods for inverse protein folding remains a cornerstone task in \textit{de novo} protein design.
We introduced a diffusion approach that fine-tunes a state-of-the-art inverse folding model to significantly accelerates the inference process.
Our work has several limitations.
First, training ProteinMPNN diffusion without pre-training performs worse than fine-tuning.
The reasons for this are unclear.
Second, our non-autoregressive generation approach does not edit previously generated residues, making it prone to error accumulation. Ideally, the model would be able to iteratively refine the whole sequence like in other non-autoregressive sequence generation works.
Third, we utilize a discrete time diffusion process that is incompatible with continuous time stochastic differential equations (SDE).
An interesting direction would to explore the continuous time formulations.
Lastly, we were not able to find correlations between purity decoding order with structural characteristics, i.e. residues close in space should decode together.
We plan to investigate further into optimal decoding orders and analyze its relation with structure.






\bibliography{reference}

\newpage
\appendix

\section{Related work}
\label{sec:related_work}
Advances in inverse folding can be broadly categorized into improving structure representation or sequence generation. On the structure side, graph-based architectures combined with geometric features led to significant improvements over MLP and CNN based models \citep{Ingraham2019, jing2021learning, gao2022pifold}. On the sequence side, Transformer-based architectures have dominated \citep{Ingraham2019, hsu2022learning, Yang2023MIF}. However, these Transformer-based architectures generate sequences autoregressively, leading to slow inference times and high variance in generation quality. Non-autoregressive alternatives have shown strong performance on benchmarks such as perplexity and sequence recovery \citep{gao2022pifold, yi2023graph}. However, these benchmarks have proved unreliable to protein designers. Independent of us, \citet{wu2022protein} found that ESM-IF \cite{hsu2022learning} achieving SOTA sequence recovery of 72\% but poor designability results compared to methods with lower sequence recovery. Multiple works have found designability has been shown to correlate with experimental validation \citep{proteinmpnn,watson2023novo, wicky2022hallucinating}. Our method improves on Protein-MPNN, a popular model among protein designers that has demonstrated strong experimentally validated results, by generating sequences faster and non-autoregressively \cite{proteinmpnn}.





\end{document}